\documentclass[fleqn,usenatbib]{mnras}

\usepackage{newtxtext}

\usepackage[T1]{fontenc}

\DeclareRobustCommand{\VAN}[3]{#2}
\let\VANthebibliography\thebibliography
\def\thebibliography{\DeclareRobustCommand{\VAN}[3]{##3}\VANthebibliography}

\usepackage{graphicx}	
\usepackage{amsmath}	
\usepackage{threeparttable}
\usepackage{caption}
\usepackage{dirtytalk}
\usepackage{amssymb}
\usepackage{ulem}
\usepackage{subcaption}
\usepackage[x11names,dvipsnames]{xcolor}
\usepackage{comment}

\newcommand{\koral}{\textsc{KORAL}}
\newcommand{\rg}{r_\mathrm{g}}
\newcommand{\tg}{t_\mathrm{g}}
\newcommand{\Mdot}{\dot{M}}

\newcommand{\pgas}{p_\mathrm{gas}}
\newcommand{\pmag}{p_\mathrm{mag}}
\newcommand{\pr}{p_\mathrm{rad}}
\newcommand{\uint}{u_\mathrm{int}}
\newcommand{\diff}{\mathrm{d}}
\newcommand{\Ledd}{L_{\mathrm{Edd}}}
\newcommand{\Medd}{\dot{M}_{\mathrm{Edd}}}
\newcommand{\gdet}{\sqrt{-g}}
\newcommand{\orcid}[1]{\href{https://orcid.org/#1}{\includegraphics[width=8pt]{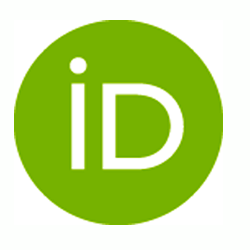}}}

\title[Radiative GRMHD of compact stars and black holes]{
Radiative GRMHD simulations of super-Eddington accretion \\on neutron stars, ultra-compact stars, and black holes }

\author[F. Kayanikhoo et al.]{
\parbox{\textwidth}{
Fatemeh Kayanikhoo$^{1,2}$\orcid{0000-0003-2835-3652}\thanks{E-mail: fatemeh.kayanikhoo@physics.slu.cz},
Debora Lan\v{c}ov\'{a}\orcid{0000-0003-0826-9787}$^{1,2}$\thanks{E-mail: debora.lancova@gmail.com},
Maciek Wielgus\orcid{0000-0002-8635-4242}$^{3}$,
Leszek Zdunik\orcid{0000-0003-4637-2160}$^{2}$, and\\
Martin Urbanec\orcid{0000-0001-9635-5495}$^{1}$
} \vspace{0.5cm}\\ 
$^{1}$Research Centre for Computational Physics and Data Processing, Institute of Physics, Silesian University in Opava, Czech Republic \\
$^{2}$Nicolaus Copernicus Astronomical Center of the Polish Academy of Sciences, Bartycka 18, 00-716 Warsaw, Poland \\
$^{3}$Instituto de Astrofísica de Andalucía-CSIC, Glorieta de la Astronomía s/n, E-18008 Granada, Spain
}

\date{Accepted XXX. Received YYY; in original form ZZZ}

\pubyear{2026}

\begin{document}
\label{firstpage}
\pagerange{\pageref{firstpage}--\pageref{lastpage}}
\maketitle

\begin{abstract}
We investigate how the compactness of an accreting object in an X-ray binary affects the accretion flow, radiative efficiency, and apparent luminosity in the super-Eddington regime. Using general relativistic radiative magnetohydrodynamic simulations, we model accretion onto canonical neutron stars, ultra-compact stars, and black holes of mass $2.2\,M_{\odot}$. In the stellar models, the disc is truncated at a magnetospheric radius that follows the standard scaling, with a proportionality factor around 0.5, close to the usually used thin-disc estimation. A smaller radius leads to a lower mass accretion rate but higher radiative and kinetic efficiencies, stronger unbound outflows and larger apparent luminosity for observers close to the axis. The most compact star cases exceed the thin-disc efficiency of a non-rotating object and reach apparent luminosities above 100 times the Eddington limit, which is well within the ultraluminous X-ray source (ULX) regime, while a larger magnetosphere weakens the beaming. An accretion disc around a black hole of the same mass is two to four times less radiatively luminous than the stellar models, even in the case of a rapidly spinning black hole, despite strong geometrical beaming.
Our results demonstrate that the presence of a stellar surface and magnetic field, together with high compactness, can produce extreme luminosities observed in ULXs and suggest that ultra-compact objects with an intrinsic magnetic field are potential accretors in ULXs.
\end{abstract}




\section{Introduction}

Ultraluminous X-ray sources (ULXs) observed in nearby galaxies are characterized by the observed luminosity exceeding $10^{39}\,\rm erg\,s^{-1}$, but reaching up to $10^{42}\,\rm erg\,s^{-1}$ in some cases, see \citet{king+2023} for a review. To explain the large observed luminosity, the ULXs were believed to host an intermediate-mass black hole (BH), with mass $M\sim10^2-10^5\,M_\odot$, where $M_\odot$ is the solar mass. 
However, X-ray pulsation was observed in several of them \citep[so-called pulsating ULXs, PULXs;][]{bachetti2014, Israel2016, Israel2017a,Townsend2017,Tsygankov2017,brightman2018, Furst2018, Carpano2018,Doroshenko2018, Heida_2019, Chandra2020}, suggesting a presence of a neutron star (NS) or other compact star accretor in at least some of the ULXs. 

The extremely high observed luminosity in the case of stellar-mass accretors, particularly those associated with PULXs, is difficult to interpret within the standard accretion disc model assuming that the luminosity is bounded by the Eddington limit, $\Ledd = (4\pi G M m_{\rm p} c)/\sigma_{\rm T} \simeq 1.257 \times10^{38}M/M_\odot \,\rm erg\,s^{-1}$, where $m_{\rm p}$ is the proton mass and $\sigma_{\rm T}$ is the Thomson scattering cross section. It may be alternatively explained by a compact star accreting in a super-Eddington regime, with the radiation beamed towards the observer along the rotational axis \citep[][]{king2001,king_masses_2009}. This model was subsequently investigated in \citet{ king2017,king2019,king2020} and we refer to it further as the KLK model. The KLK model was tested in numerical simulations of accreting NSs, where the apparent luminosity reached more than $100\,\Ledd$ \citep[][]{2018MNRAS.479.3936A,2021ApJ...917L..31A,kayanikhoo2025a, kayanikhoo2025b}.
An alternative class of models involves a very strong surface magnetic field ($10^{13}$--$10^{14}\,$G) resulting in reduced scattering opacity in the accretion columns and thus shifting the Eddington limit \citep{2015MNRAS.448L..40E, 2015MNRAS.449.2144D, 2015MNRAS.454.2539M}. These two models differ mainly in the assumed magnetic field strength and in the collimation of the escaping radiation. In this work, we simulate stars with dipole fields of $10^{10}$--$10^{11}\,$G in the KLK scenario; the strong-field regime is currently not possible to study with our setup, since the corresponding magnetospheric radius would exceed the size of the simulated disc.

Beyond the canonical NSs, even more compact stars may in principle power the ULXs. The characteristics of such objects are still poorly understood, particularly in the regime of extreme compactness that may be associated with exotic equations of state \citep[EOS, e.g.,][]{Oertel2017}. A higher compactness of a similar mass star deepens the gravitational potential well and impacts the structure of the inner accretion flow. This, in turn, modifies the coupling between the stellar magnetic field and the disc, the geometry of the accretion columns, as well as the efficiency and collimation of the radiation-driven outflows. In the context of ULXs, such changes in accretion geometry and beamed radiation can significantly affect the apparent luminosity and the angular dependence of the emitted flux. This motivates a systematic exploration of how the compactness of the accretor shapes the accretion flow and the resulting observables. In \cite{kayanikhoo2025b} we investigated how the accretion rate and the NS magnetic field strength affect the accretion efficiency and geometrical beaming. We showed that a weaker dipole field leads to a higher outflow rate and greater radiative efficiency, resulting in brighter ULXs. We also found that as the accretion rate increases, radiation energy is more efficiently converted into the kinetic energy of the outflow; nevertheless, strong geometrical beaming still results in high apparent luminosities.

In this work, we use numerical general relativistic radiative magnetohydrodynamics (GRRMHD) to examine how the compactness of the central object changes the properties of the accreting system, focusing on the observables relevant in connection to the ULXs. We fix the central object mass to $M_*=2.2\,M_\odot$ and vary its radius, thus affecting the compactness. We consider static stellar objects with a surface radius $R_*$ larger than or equal to $2.3\,\rg$, where $\rg=GM/c^2$, as well as both static and rotating BHs with dimensionless spin parameter ranging from $a=0$ to $a=0.997$. For the star cases, we model a dipolar magnetic field configuration. We assume the same initial distribution of matter around the central object for all simulations and vary the magnetic field configuration. 

This work is organized as follows: In Section~\ref{sec:koral}, we introduce the \koral{} code and the numerical methods used here, and in Section~\ref{sec:init} we comment on the initial state and simulation parameters. The formulae used to compute the relevant physical quantities are given in Section~\ref{sec:measured_quantities}. The results are presented in Section~\ref{sec:res}, and the conclusions and discussions are given in Section~\ref{sec:conc}.

\section{Numerical methods}\label{sec:koral}

We use a GRRMHD code \koral{} \citep{Sadowski2013,Sadowski2014} with modifications enabling us to simulate accretion onto NSs \citep{2018MNRAS.479.3936A,2021ApJ...917L..31A}. \koral{} 
solves the co-evolution of the fluid, magnetic field, and radiation field in a stationary spacetime given by the metric tensor $g_{\mu\nu}$ on a fixed grid.

\subsection{Radiative GRMHD equations}

 The code solves the equations of fluid continuity, energy and momentum conservation, and the source-free Maxwell equation under the ideal MHD approximation,
\begin{align}
&\nabla_\mu \left(\rho u^\mu\right) = 0, \\
&\nabla_\mu \left( T^{\mu\nu} + R^{\mu\nu} \right) = 0, \\
&\nabla_\mu \left( b^\mu u^\nu - b^\nu u^\mu   \right) = 0,\label{eq:conservations}
\end{align}

\noindent where $\rho$ is the fluid rest-mass density, $u^\mu$ is its four-velocity, $T^{\mu\nu}$ and $R^{\mu\nu}$ are the MHD and radiation stress-energy tensors, respectively. Four-vector $b^\mu$ is the fluid-frame magnetic field, related to the coordinate-frame $B^i$ through \citep{1999MNRAS.303..343K}
\begin{align}
b^t &= B^i u^\mu g_{i\mu}, \\
b^i &= \dfrac{B^i + b^t u^i}{u^t} \ .
\end{align}
Then we have the MHD stress-energy tensor
\begin{equation}
T^{\mu}\phantom{}_\nu= \left(\rho + u_{\mathrm{int}} + \pgas + b^2\right)u^\mu u_\nu + 
\left(\pgas+\dfrac{1}{2}b^2\right)\delta^{\mu}\phantom{}_\nu - b^\mu b_\nu,
\label{eq:Tmunu}
\end{equation}

\noindent where $b^2 = b^\mu b_\mu$ and $\uint$ is the gas internal energy, which, under the ideal gas assumption, is related to the gas pressure $\pgas$ as 
\begin{equation}
    \uint = \frac{\pgas}{\Gamma - 1},
    \label{eq:eos}
\end{equation}
\noindent with the assumed adiabatic index $\Gamma = 5/3$.

The radiation tensor $R^{\mu\nu}$ is found using the M1 closure scheme \citep{1984JQSRT..31..149L,Sadowski2014}, which assumes the existence of a \say{radiation frame} $u_\mathrm{R}^\mu$ in which the radiation is isotropic. The radiation tensor is then 
\begin{equation}
    R^{\mu\nu} = \dfrac{4}{3}\bar{E} u_{\mathrm{R}}^\mu u_{\mathrm{R}}^\nu + \dfrac{1}{3}
    \bar{E} g^{\mu \nu},
    \label{eq:Rmunu}
\end{equation}
\noindent where $\bar{E}$ is the radiation energy density in the radiation frame. The gas and radiation exchange energy and momentum via the radiation four-force density $G^\nu$,
\begin{equation}
\widehat{G}^\nu = \widehat{G}_\mathrm{Compt}^\nu + \widehat{G}_0^\nu,
\end{equation}

\noindent where the overhat denotes quantities evaluated in the fluid frame, and which includes a term corresponding to thermal Comptonization $\widehat{G}_\mathrm{Compt}^\nu$ \citep{2015MNRAS.454.2372S}, as well as

\begin{equation}
\widehat{G}^\nu_0 = 
    \begin{pmatrix}
    \kappa_\mathrm{a}\rho\left(\widehat{E} - 4\pi \widehat{B}\right)\\
    \left(\kappa_\mathrm{a} + \kappa_\mathrm{es}\right)\rho\widehat{F}^i
    \end{pmatrix},
\end{equation}

\noindent where $\kappa_\mathrm{a} =  6.4\times10^{22} \left(\rho/\rm g\,cm^{-3}\right) \left(T_{\rm g}/\rm K\right)^{-7/2}\ \mathrm{cm^2\,g^{-1}}$ is the free-free absorption for gas temperature $T_{\rm g}$, and $\kappa_\mathrm{es} = 0.34\,{\rm cm^2g^{-1}}$ is the scattering opacity, $\widehat{B}$ is the frequency-integrated Planck function and $\widehat{F}^i$ are the spatial components of the radiation flux.

\subsection{Force-free electrodynamics scheme}

In the innermost region surrounding the central star, where the gas magnetization $\sigma = b^2/\rho$ is extremely high, the force-free electrodynamics (FF) is used \citep{2017ApJ...851L..34P,2021ApJ...917L..31A}, which allows us to resolve the highly magnetized low-density area in the magnetosphere of the central star without the artificial injection of gas, related to numerical floors enforced in standard GRMHD. In our simulations, the area where FF is employed is fixed to $r_{\mathrm{FF}} \leq 20\,\rg$. Recently, a hybrid FF solver was implemented in the \koral{} code, allowing for a smooth transition between MHD and FF solutions \citep{2024MNRAS.532.3198C}, but only for non-radiative MHD. However, in our case, the star magnetosphere is expected to be approximately spherical, and the area of high $\sigma$ is well-defined; thus the fixed-FF zone is appropriate. 

We follow the FF formulation of \cite{dabarca_thesis}. The FF scheme employs a static density floor $\rho_f$ following a broken power-law profile in radius, chosen such that the magnetization peaks at $\sigma \sim 10^4$ on the NS surface, while at larger radii 
$\rho_f \propto r^{-6}$, maintaining an approximately constant magnetization. 
The contribution of this floor to the total density is tracked via an additional 
scalar field $\mathcal{F}$, evolved as
\begin{equation}
    \nabla_{\mu}\left(\mathcal{F}\,\rho\, u^{\mu}\right) = 0,
\end{equation}
where $\mathcal{F}$ represents the fraction of the local density originating from the floor. If $\rho < \rho_f$, the density is brought up to $\rho_f$ and $\mathcal{F}$ is set to $(\rho_f - \rho)/\rho_f$; if the floor contribution grows too large $\left(\mathcal{F}\rho > \rho_f\right)$, it is partially subtracted. The internal energy is treated analogously through a ceiling value 
$\mathcal{U}$: when $u_{\mathrm{int}} > \mathcal{U}$, it is linearly interpolated toward $\mathcal{U} + \left(1-\mathcal{F}\right)
\left(u_{\mathrm{int}} - \mathcal{U}\right)$.

The fluid four-velocity is subsequently corrected by splitting it into components parallel and perpendicular to the magnetic field measured by a static observer with four-velocity $u^{\mu}_{\rm st} = (u^t_{\rm st}, 0, 0, 0)$,
\begin{equation}
    u^{\mu}_{\parallel} = k\,\frac{b^{\mu}_{s}}{b^{2}_{s}}, 
    \qquad 
    u^{\mu}_{\perp} = u^{\mu} - k\,\frac{b^{\mu}_{s}}{b^{2}_{s}},
\end{equation}
where $k \equiv u_{\mu} b^{\mu}_{s}$ and 
$b^{\nu}_{s}$ is the magnetic field four-vector evaluated in the stationary observer frame. 
The four-velocity of the fluid is then replaced by $A\left(u^{\mu}_{\perp} 
    + \mathcal{F}\,u^{\mu}_{\parallel}\right)$,
%
with $A = {\left(u^{2}_{\perp} + \mathcal{F}^{2}\,u^{2}_{\parallel}\right)}^{-1/2}$. The parallel component, which is prone to unphysical growth at high magnetizations, is thereby suppressed, while the perpendicular component essential for the evolution of the magnetic field in the GRRMHD equations is retained. This correction is applied after each conserved-to-primitive inversion step.

\section{Initialization and boundary conditions}
\label{sec:init}

We performed a large set of two-dimensional axisymmetric simulations of accretion onto a central object with various compactness. The simulations are split into two categories according to the nature of the central object: compact stars with varying radius (and compactness) and surface magnetic field strength, and BHs for which we use the Kerr metric and consider spins $a=(0.0, 0.3, 0.6, 0.997)$. The first group of models corresponds to a non-rotating central object, whose surrounding is described by the exterior Schwarzschild metric. The stellar surface, located at $R_*$ with a dipolar magnetic field geometry, is treated as a boundary condition \citep[see][for the details on the numerical treatment of the star surface]{2021ApJ...917L..31A}.

In all cases, the initial torus is based on the axisymmetric stationary solution of \citet{2013A&A...559A.116P}, including both gas and radiation contribution to the total pressure. As the initial torus is radiation-pressure dominated, we initialize it using $\Gamma_0 = 4/3$. The torus parameters are summarized in Table~\ref{tab:torus} and are the same for all simulations. We keep the initial plasma $\beta_0$ in the torus centre the same for all cases.  The magnetic field strength at the torus centre is generally smaller than $0.3\%$ of the star surface magnetic field strength.  
The main differences in the simulations are the initial magnetic field topology and the star radius. In all simulations, the initial torus magnetic field topology is chosen such that the standard and normal evolution \citep[SANE;][]{narayanGRMHDSimulationsMagnetized2012} state develops. In all simulations, we employ an artificial mean-field dynamo term, as introduced in \citet{sadowski2015}, in order to sustain the magnetorotational instability (MRI) in long-running axisymmetric simulations.

\begin{table}
\begin{threeparttable}
    \caption{Initial torus parameters and their meaning \citep[see][]{2013A&A...559A.116P}.}
    \label{tab:torus}
    \centering
    \begin{tabular}{c|c|c}
        \hline
         Parameter & Value & Meaning  \\
         \hline \hline 
         $r_1$ & 45 $\rg$ & Inner angular momentum break radius \\
         $r_2$ & 200 $\rg$ & Outer angular momentum break radius \\
         $\zeta$ & 0.995 & Keplerian angular momentum fraction \\
         $\mathcal{K}$ & $11\times10^2$ & Entropy constant (in code units)\\ 
         $\Gamma_{0}$ & $4/3$ &Adiabatic index \\ 
         $\beta_0$ & $20$ & Pressure ratio $(\pgas+\pr)/\pmag$\\ 
         $B_{\mathrm{torus}}$ & $3\times10^{7}$ & Magnetic field in torus centre in G \\ \hline
    \end{tabular}

\end{threeparttable}
\end{table}

\begin{table*}
\begin{threeparttable}
    \caption{List of the simulations with labels and their parameters, the total duration $\Delta t_\mathrm{tot}$ and the length of the averaging window $\Delta t_\mathrm{avg}$. Selected accretion flow properties are listed after: the net mass accretion rate $\Mdot$ and unbound outflowing mass accretion rate $\Mdot^{\rm ub}_{\rm out}$, apparent luminosity $L_{\rm iso}$, radiative efficiency $\eta_{\rm rad}$ and total efficiency $\eta_{\rm tot}$ and the inverse beaming factor $1/b$.}
    \label{tab:sims}
    \centering
    \begin{tabular}{c c c c c c c c c c c c c}
        \hline
Simulation & $R_*$  & $B_*$  & $B\,\left(6\,\rg \right)$ & $\mu$ & $\Delta t_\mathrm{tot}$ & $\Delta t_\mathrm{avg}$ & $\dot M$ & $-\dot M_{\rm out}^{\rm ub}$ & $L_\mathrm{iso}$  & $\eta_\mathrm{rad}$ & $\eta_\mathrm{tot}$ & $1/b$ \\
& $\left(\rg\right)$ & $\left(10^{10}\,\rm G\right)$ & $\left(10^{10}\,\rm G\right)$ & $ \left( \mu_{\rm FDR} \right)$ &$\left(\tg\right)$ & $\left(\tg\right)$ & $\left(\dot{M}_{\mathrm{Edd}}\right)$ & $\left(\dot{M}_{\mathrm{Edd}}\right)$ & $\left(L_{\mathrm{Edd}}\right)$ & -- & -- &--\\\hline \hline
\texttt{FDR-R2.3} & 2.3  & 58.20 & 1.22 & 1 & 60 000 & 44 000 & 5.16 & 46.7 & 181.4 & 0.141 & 0.249 & 25.1 \\ \hline
\texttt{FDR-R2.5} & 2.5  & 36.10 & 1.22 & 1 & 60 000 & 44 000 & 9.74 & 37.4 & 143.5 & 0.065 & 0.105 & 23.0 \\ \hline
\texttt{FDR-R3.2} & 3.2  & 11.84 & 1.22 & 1 & 60 000 & 44 000 & 17.56 & 19.1 & 86.5 & 0.027 & 0.035 & 18.6 \\ \hline
\texttt{FDR-R4.0} & 4.0  & 5.02 & 1.22 & 1 & 60 000 & 44 000 & 19.29 & 19.3 & 73.8 & 0.021 & 0.025 & 18.8 \\ \hline
\texttt{FDR-R5.0} & 5.0  & 2.27 & 1.22 & 1 & 60 000 & 44 000 & 19.39 & 22.3 & 74.3 & 0.021 & 0.026 & 18.3 \\ \hline
\texttt{FDR-R6.0} & 6.0  & 1.22 & 1.22 & 1 & 60 000 & 44 000 & 19.83 & 16.0 & 76.3 & 0.022 & 0.028 & 17.7 \\ \hline
\texttt{SD3-R2.5} & 2.5  & 3.92 & 0.13 & 0.11 & 35 000 & 25 000 & 17.10 & 18.7 & 102.5 & 0.032 & 0.042 & 19.0 \\ \hline
\texttt{SD3-R3.2} & 3.2  & 2.69 & 0.28 & 0.23 & 53 000 & 35 000 & 18.73 & 25.8 & 93.7 & 0.026 & 0.033 & 19.6 \\ \hline
\texttt{SD3-R4.0} & 4.0  & 2.33 & 0.54 & 0.44 & 53 000 & 35 000 & 17.40 & 19.1 & 81.6 & 0.029 & 0.034 & 16.5 \\ \hline
\texttt{SD3-R5.0} & 5.0  & 1.97 & 1.06 & 0.87 & 53 000 & 35 000 & 19.65 & 18.6 & 74.8 & 0.022 & 0.026 & 17.7 \\ \hline
\texttt{SD7-R3.2} & 3.2  & 6.28 & 0.65 & 0.53 & 60 000 & 44 000 & 21.46 & 19.7 & 72.5 & 0.026 & 0.028 & 13.2 \\ \hline
\texttt{SD7-R6.0} & 6.0  & 4.28 & 4.28 & 3.50 & 60 000 & 44 000 & 27.30 & 12.9 & 63.6 & 0.023 & 0.025 & 10.2 \\ \hline
\texttt{BHa00} & 2.0\,\tnote{a}  & -- & -- & -- & 110 000 & 49 000 & 10.47 & 3.4 & 24.9 & 0.018 & 0.022 & 13.2 \\ \hline
\texttt{BHa03} & 1.95\,\tnote{a}  & -- & -- & -- & 91 000 & 53 000 & 8.47 & 4.5 & 27.3 & 0.023 & 0.030 & 14.3 \\ \hline
\texttt{BHa06} & 1.80\,\tnote{a}  & -- & -- & -- & 130 000 & 50 000 & 7.38 & 12.0 & 42.3 & 0.026 & 0.044 & 21.9 \\ \hline
\texttt{BHa0997} & 1.08\,\tnote{a}  & -- & -- & -- & 115 000 & 33 000 & 5.51 & 9.2 & 60.0 & 0.052 & 0.084 & 20.9 \\ \hline\hline

    \end{tabular} 
    \begin{tablenotes}
        \item[a] radius of the BH event horizon.
    \end{tablenotes}
\end{threeparttable}
\end{table*}

\subsection{Compact star simulations}

\begin{figure}
    \centering
    \includegraphics[width=1\linewidth]{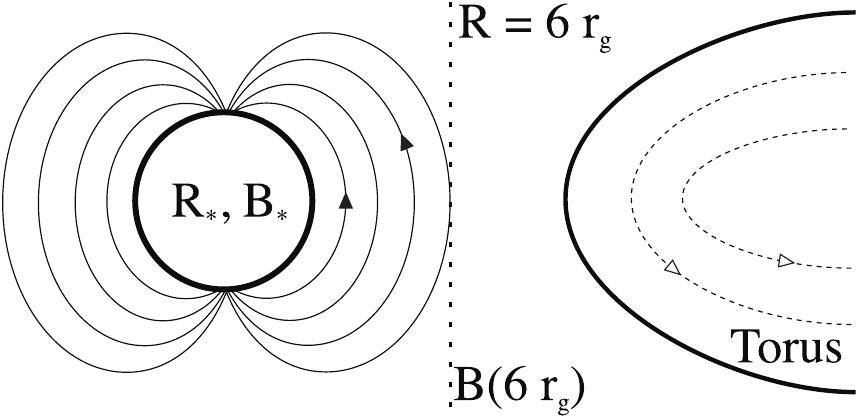}
    \caption{Schematic illustration of the star-torus configurations. The solid bold lines show the star surface and the torus outer surface. $R_*$ and $B_*$ denote the star radius and surface magnetic field, respectively. The solid thin lines show the magnetic dipole. The dashed vertical line represents radius $R = 6\,r_{\rm g}$ at which the magnetic field strength is fixed to $1.22\times 10^{10}\,$G for the \texttt{FDR} series. For the \texttt{SDs} series, this value varies depending on the stellar radius; see column $B(6 \,r_{\rm g})$ in Table~\ref{tab:sims}. The torus magnetic field is shown with the dotted lines. The figure is not to scale.}
    \label{fig:models}
\end{figure}

In the compact star cases, the torus magnetic field is initially threaded with magnetic field loops following the equipressure surfaces. The central star of radius $R_*$ is initialized with a relativistic magnetic dipole moment $\mu$ \citep{1983ApJ...265.1036W} oriented oppositely to the torus field and aligned with the torus rotational axis. The value of $\mu$ fixes the magnetic field strength. A schematic of this configuration is shown in Fig.~\ref{fig:models}.

The torus and the central star are embedded in an initially static background atmosphere with density and internal energy lower than those of a torus by several orders of magnitude, declining towards the outer boundary as a broken power-law. This creates a hot atmosphere around the central star and a colder one around the torus. 

 We perform the star simulations with the resolution $512\times510$, employing logarithmic spacing of grid cells in the radial direction. This approach provides higher resolution at radii closer to the star surface, which is essential for resolving the structure of the accretion columns. The resolution remains sufficient to resolve the MRI even at large radii. Two grid cells above the star surface are designated as a boundary layer, where kinetic and internal energy are converted into radiation, part of which is reflected from the surface. In the simulations presented in this study, the albedo is set to 75\% \citep{2021ApJ...917L..31A,kayanikhoo2025b}. The polar-angle domain in principle extends from $0$ to $\pi$; however, to avoid numerical issues near the axis, we exclude a narrow region around each pole by restricting the polar coordinate to $(\theta_{\rm min},\pi - \theta_{\rm min})$ where $\theta_{\rm min} = 1.5^\circ$. The magnetic field at the star surface $B_*$ is a boundary condition, kept at its initial value throughout the simulation.

We performed 12 simulations of accretion onto a compact star, varying the star radius $R_*$ (and thus the compactness $M/R_*$) and the surface magnetic field strength. The complete set of models is summarized in Table~\ref{tab:sims}. The models are divided into two series: the fixed dipole radius (\texttt{FDRs}) and surface dipole (\texttt{SDs}), based on the configuration of the magnetic field on the star surface. 

For the first series, \texttt{FDR-R\#}, we initialize the dipolar magnetic field of the star in such a way that the dipole moment $\mu = \mu_{\rm FDR}$ is the same for all the simulations in this sequence, regardless of the compactness of the central object. The chosen dipole moment results in a magnetic field strength measured by a static observer at $(r,\theta) = (6\,r_{\rm g}, 0)$ is $1.22 \times 10^{10}$\,G, fixed for all \texttt{FDR-R\#} simulations, see Table~\ref{tab:sims}. This prescription fixes the magnetic field at the star surface (different for each simulation), that we subsequently keep as a boundary condition.

The other series of models, \texttt{SDs}, includes \texttt{SD3–R\#} and \texttt{SD7–R\#}. There, we fix the quantity $\mu R_*^{-3}$, which technically corresponds to the magnetic field strength that would be inferred at a star surface $R_*$ for a flat-spacetime dipole. However, because of the relativistic corrections \citep{1983ApJ...265.1036W} the magnetic field measured by a static observer on the star surface varies within a factor of 2 with $R_*$, see Table~\ref{tab:sims} for the exact values of $B_*$. These values are preserved as a boundary condition of a corresponding simulation. For a given $R_*$, at a fixed radial coordinate, magnetic field in a model \texttt{SD7–R\#} is 7/3 times stronger than that of a corresponding model \texttt{SD3–R\#}. The notation \texttt{R\#} (e.g., \texttt{R2.3, R2.5, R4.0}) indicates the star radius $R_*$.

\subsubsection*{Minimum compact-star radius}

For a causal-limit EOS, where the sound speed $v_{\rm s}$ does not exceed the speed of light $c$, the minimum radius of a non-rotating isotropic compact star is $R_*^{\rm min}\simeq 2.82\, \rg$ \citep[see section 6.6.4 in][]{neutron-stars-1}. If this condition ($v_{\rm s}\lesssim c$) is relaxed, the star radius limit is, for an incompressible fluid corresponding to the Buchdahl limit, $R_*^{\rm min} = 2.25\,\rg$ \citep{Buchdahl1959}. For a realistic microphysical EOS, the minimum value of $R_*^\mathrm{min} \approx 3.3\,\rg$ for $M_{\rm max}=2.05 \,M_\odot$ in the case of the SLy model \citep{Douchin2001}, or $R_*^\mathrm{min} \approx 3.25\,\rg$ for $M_{\rm max}=2.78\, M_\odot$ in the case of the NL3 model \citep{PhysRevC.55.540}. The recent study by \cite{Rezzola2026} indicates that the minimum stellar radius is $R^\mathrm{min}_* =3 \,\rg$, and this limit is attained by the maximum-mass configuration. Our models with $ R_* \geq 3.2\,\rg$ (canonical NS) are well within this constraint. We also consider more compact models with minimum radius of $R_* = 2.3\,\rg$ as theoretical ultra-compact objects, possibly involving exotic microphysics, which serves as a transition case toward the BH models and satisfies the Buchdahl limit.

\begin{figure*}
    \centering
    \includegraphics[width=1\linewidth]{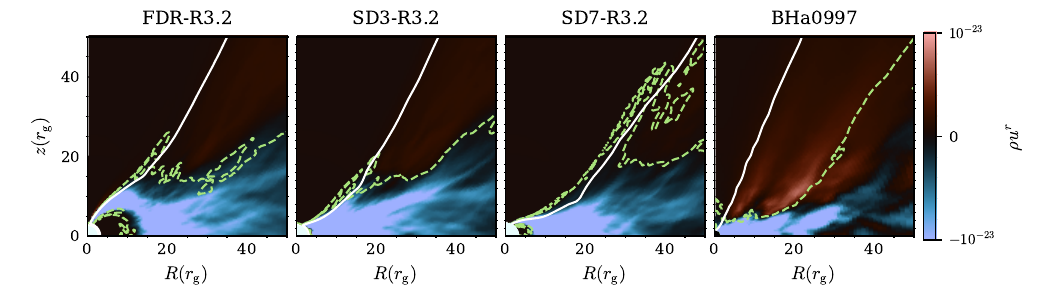}
    \caption{The gas momentum density $\rho\,u^r$ in code units shown for a representative simulation from each category. Negative values indicate the infalling material and the positive values are the outflowing material. The dashed green line is the $\xi = 0$ surface (Eq.~\ref{eq:mu}), which separates the bound and unbound regions. The white solid line is the photosphere $\tau=1$ (Eq.~\ref{eq:tau}). A region of optically thick outflows (between the white and green lines) develops in all cases, although only at larger radii in case of \texttt{SD7-R3.2}.}
    \label{fig:momentum}
\end{figure*}

\subsection{Black hole simulations}

We additionally performed a set of simulations of accretion on a non-rotating, mildly, and fast rotating BH. In this case, there is no dipolar magnetic field of the central BH enforced by a boundary condition \citep[see also][]{2021AN....342..357K, 2024ApJ...968L..10S,2026ApJ...999..121P}. 
We set up the initial conditions as close as possible to the star simulation, with the same torus and central object mass. The simulations are labelled as \texttt{BHa\#}, where 
\texttt{a\#} denotes the BH spin.

Initializing the torus with the same poloidal field as in the star case resulted in large magnetic flux on the BH horizon, thus the disc became magnetically arrested \citep[MAD; ][]{2003PASJ...55L..69N}. In order to avoid reaching MAD state, we initialized the BH simulations with multiple magnetic loops of opposite polarity inside the torus while keeping the same pressure ratio $\beta_0$. Although this setup led to a relatively high horizon magnetic flux, it remained well below the MAD limit throughout the whole run \citep[][]{2011MNRAS.418L..79T}.

The BH simulations were performed with a resolution of 256$\times$256, using a grid that is more concentrated toward the central equatorial zone, since there was no need to resolve the accretion columns far from the equatorial plane. In the absence of an extended magnetosphere it generally takes more time for a BH simulation to relax to a quasi-stationary state. On the other hand, the duration for which the artificial dynamo term can maintain the turbulent angular momentum transport in the axisymmetric setup is limited to approximately 150 000 $\tg$, where $\tg = GM/c^3$ \citep{sadowski2015}. Nonetheless, we were able to obtain comparable durations of analysed simulation segments for all cases, see Tab.~\ref{tab:sims}.

\begin{figure*}
          \includegraphics[width=\linewidth]{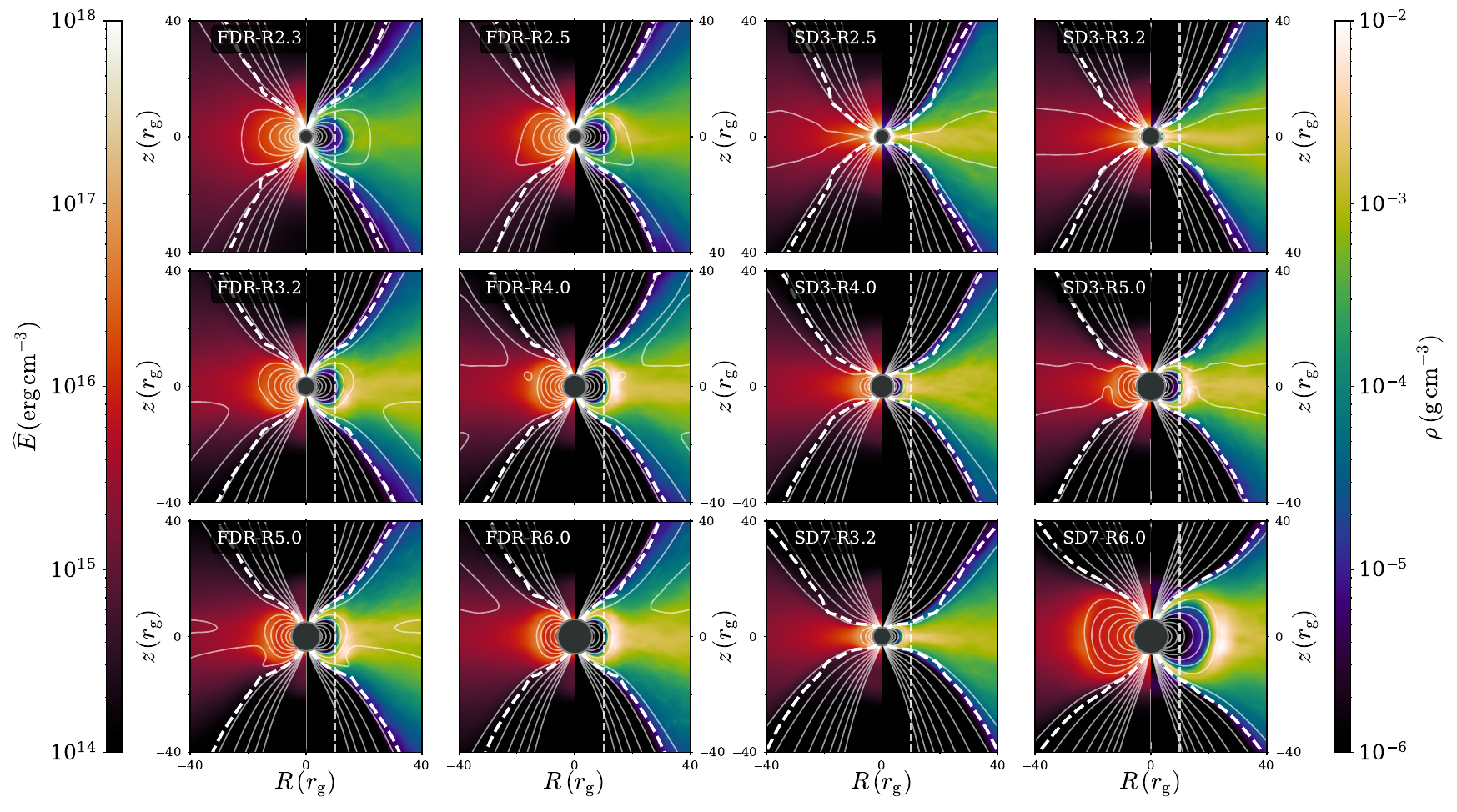}
   \caption{Maps of the rest-mass density $\rho$ (right half) and radiation energy density $\hat{E}$ (left half) from the time-averaged data of the star simulations are shown; the simulation models are labelled. The thin white contours indicate the magnetic field lines, the bold white dashed contour shows the photosphere $\tau = 1$ (see Eq.~\ref{eq:tau}) and the vertical dashed line marks $R = 10\,\rg$ as a guideline.}
    \label{fig:avg_grid}
\end{figure*}

\section{Measured quantities}
\label{sec:measured_quantities}

In this section, we present definitions of all quantities that we measure from our simulations to make comparisons between the models.

\subsection{Photosphere and inflow/outflow regions}

The optical depth $\tau (r,\theta)$ computed with respect to electron scattering, which dominates throughout the domain, is calculated as 
\begin{equation}
\tau \left(r,\theta\right) = \int_{0}^{\theta} \rho \left(r,\theta'\right) \kappa_{\mathrm{es}} r \diff \theta'.
\label{eq:tau}
\end{equation}
The photosphere is located where $\tau(r,\theta)=1$.

To distinguish bound from unbound gas in the presence of radiation and strong poloidal fluxes in the accretion columns, we have to consider a generalized Bernoulli parameter. We combine the approaches of \citet{narayanGRMHDSimulationsMagnetized2012,2013MNRAS.436.3856S} and \citet{curdGRRMHDSimulationsMAD2023} and define a normalized poloidal energy-flow rate,
\begin{equation}\label{eq:mu}
    \xi =  -\frac{(T^{\mathrm{P}}_{\ t} + R^{\mathrm{P}}_{\ t})}{\rho\, u^{\mathrm{P}}} - 1 ,
\end{equation}
\noindent where the superscript $P$ denotes the poloidal projection and the subtraction of unity removes the contribution of the rest mass. Gas is then considered unbound where $\xi > 0$. The poloidal projection is constructed from the orthonormal radial and polar components of the four-velocity,
\begin{equation}
    u^{\widehat r} = \sqrt{g_{rr}}\,u^{r}, \qquad
    u^{\widehat \theta} = \sqrt{g_{\theta\theta}}\,u^{\theta},
\end{equation}
\noindent with the poloidal velocity given by $u^\mathrm{P} = \sqrt{(u^{\widehat r})^2 + (u^{\widehat \theta})^2}$. The stress-energy tensor component $T^\mathrm{P}_{\ t}$ is then obtained as a projection 
\begin{equation}
    T^\mathrm{P}_{\ t} = \sqrt{g_{rr}}\,T^{r}_{\ t}\,\widehat n^r + \sqrt{g_{\theta\theta}}\,T^{\theta}_{\ t}\,\widehat n^\theta,
\end{equation}
where $\widehat{n}^r = u^{\widehat r}/u^\mathrm{P}$ and $\widehat{n}^\theta = u^{\widehat \theta}/u^\mathrm{P}$. The radiation tensor poloidal projection $R^{\mathrm{P}}_{\ t}$ is obtained in the same way.

Fig.~\ref{fig:momentum} gives an illustration of the $\xi=0$ contours for selected simulations from each category. 
It is apparent that the $\xi$ parameter, as defined here, captures the strongly bound inflow in the accretion column as well as the optically thick winds on the disc surface.

\subsection{Mass accretion rate}

The net mass accretion rate is defined as 
\begin{equation}
\dot{M}\left(r\right) = -2\pi\int_0^\pi\gdet \rho u^r\diff \theta,
\end{equation}\label{eq:acc}
\noindent where $g$ is the metric determinant. We give values of $\Mdot$ in the units of $\Medd = \Ledd/\eta c^2$, where $\eta$ is the accretion efficiency which we fix to $\eta = 0.1$ for the $\Medd$ unit definition. In Table~\ref{tab:sims} we give $\Mdot$ computed at the horizon in the case of the BH simulations. In the compact star cases we give $\Mdot$ measured at $r = 22\,\rg$, which is inside the accretion disc, below the inner edge of the initial torus, and outside of the magnetospheric radius for all presented cases. In a converged quasi-stationary simulation this net mass accretion rate $\Mdot$ does not vary much with radius. We confirm that is the case for our simulations of compact stars. 

We further define the unbound mass flow rate, which is calculated by integrating the mass flux in those parts of the domain where the radial velocity is positive and $\xi > 0$, 
\begin{equation}
\dot{M}^{\rm ub}_{\rm out}\left(r\right) =  -2\pi\int_0^\pi\gdet \rho u^r\diff \theta \bigg|_{\xi>0,\, u^r>0},
\end{equation}
and the values in Table~\ref{tab:sims} are measured at $r=200\,\rg$.

\subsection{Luminosities and efficiencies}

We calculate the bolometric radiative luminosity by integrating the radial radiative flux in the optically thin part of the domain,
\begin{equation}\label{eq:lbol}
    L_\mathrm{bol}\left(r\right) = -2\pi \int^\pi_0 \gdet R^r\phantom{}_t(\theta) \bigg|_{\tau < 1} \diff \theta.
\end{equation}
The kinetic luminosity is computed by integrating the outflowing and unbound mass flux of the gas,
\begin{equation}\label{eq:kinenergylum}
    L_{\rm kin} (r) = - 2\pi \int^\pi_0 \sqrt{-g} \left(u_t + \sqrt{-g_{tt}}\right)\rho u^r   \diff{\theta}\bigg|_{\xi>0,\, u^r>0} .
\end{equation}

\noindent The isotropic (apparent) luminosity $L_{\rm iso}(\theta)$ is defined as 
\begin{equation}\label{eq:liso}
    L_{\rm iso} (\theta) = - 4 \pi d^2 R^r\phantom{}_t (\theta)\bigg|_{\tau<1} \ .
\end{equation}
We measure the radiative flux $R^r\phantom{}_t (\theta)$ at a distance $d = 600\,r_{\rm g}$, where the radiation is expected to leave the system towards infinity without any obstruction. The beaming factor is then
\begin{equation}\label{eq:beam}
    b(\theta) = \frac{L_{\rm bol}}{L_{\rm iso}(\theta)}.
\end{equation}
Furthermore, we define the radiative and kinetic efficiencies
\begin{align}\label{eq:etarad}
   \eta_\mathrm{rad} = \frac{1}{\Mdot}L_\mathrm{bol}\left(600\,\rg \right),\\
    \eta_\mathrm{kin} = \frac{1}{\Mdot}L_\mathrm{kin}\left(600\,\rg\right),
\end{align}
and the total efficiency is $\eta_\mathrm{tot} = \eta_\mathrm{rad}+\eta_\mathrm{kin}$.

\section{Simulation results}\label{sec:res}

We present a set of 12 GRRMHD simulations with a central magnetized star of a fixed mass, varying compactness (radius) and magnetic field strength, and 4 simulations with a central BH of varying spin. The results are summarized in Table~\ref{tab:sims} and in Fig.~\ref{fig:parameters}\footnote{The results can be also plotted using tool available at \href{https://codeberg.org/dlancova/CompactnessProjectSimeExplorer/}{https://codeberg.org/dlancova/CompactnessProjectSimeExplorer/}}. All shown results are averaged over a time window $\Delta t_{\mathrm{avg}}$ starting after the initial relaxation when the mass accretion rate stabilises. Fig.~\ref{fig:avg_grid} illustrates the overall morphology of the obtained numerical solutions, showing gas density and radiation energy density maps for all star simulations, together with the photosphere and magnetic field lines. Throughout the paper, ($r,\theta$) denotes the spherical coordinates, while ($R,z$) the cylindrical ones.
\subsection{Magnetospheric radius}
\begin{figure*}
    \centering
    \includegraphics[width=\linewidth]{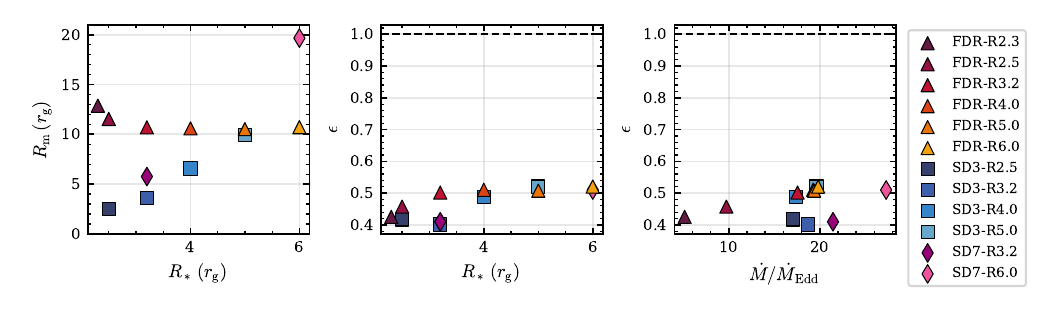}
    \caption{\textit{Left:} Disc truncation / magnetospheric radius $R_{\rm m}$. \textit{Middle:} Ratio of magnetospheric to Alfvén radius $\epsilon = R_{\rm m}/R_{\rm A}$ as a function of the star radius $R_*$. \textit{Right:} $\epsilon$ as a function of the mass accretion rate $\Mdot/\Mdot_{\rm Edd}$.
    }
    \label{fig:alfven}
\end{figure*}

The magnetic field lines from the star and from the torus interact and reconnect quickly due to their opposite initial orientation, facilitating accretion. At a magnetospheric radius $R_{\rm m}$, the pressure from the dipolar magnetic field of the star balances the pressure of the accreting plasma, resulting in a truncation of the accretion disc \citep{Elsner1977}. In the region dominated by the dipolar magnetic field, column accretion develops, with plasma guided toward the star surface by the magnetic field lines outside of the equatorial plane. The detailed geometry of the column is determined by the strength of the stellar magnetic field, radius of the compact star, and the properties of the accretion flow. Fig.~\ref{fig:avg_grid} illustrates the accretion disc/column geometry and the differences of the magnetospheric radius $R_{\rm m}$ between the models.

Balancing magnetic energy density with the kinetic energy density gives a value equal to the Alfvén radius $R_{\rm A}$. For a spherical free-fall we obtain a common proxy of the $R_{\rm m}$\footnote{Following this definition $R_{\rm A}$ is larger than the equilibrium radius calculated by \citet{Elsner1977} based on the pressure balance by a factor of $4^{1/7} \approx 1.219$. },
\begin{equation}
R_{\rm A} = \left(\frac{\mu^4}{2GM_*\Mdot^2}\right)^{1/7} \approx \ R_{\rm m} \ .
\label{eq:RAlf}
\end{equation}
In accretion discs, where thermal pressure often dominates over ram pressure, a proportionality between $R_{\rm m}$ and $R_{\rm A}$ is expected, 
\begin{equation}
 R_{\rm m} =  \epsilon R_{\rm A} \ ,
\end{equation}
with $\epsilon \approx 0.52$ estimated for a thin disc case \citep{Ghosh1979a,Ghosh1979b}.
In the left panel of Fig.~\ref{fig:alfven} we show the magnetospheric radius (measured from the simulation as the location of the disc truncation) as a function of $R_*$. The corresponding values of $\epsilon$ are shown in the middle and right panels of Fig.~\ref{fig:alfven}. For the \texttt{FDR} sequence we find $R_{\rm m} \approx 10\,\rg$, only mildly increasing toward smaller $R_*$. This is broadly consistent with expectations from a constant $\mu$ setup.

We measured values of $\epsilon$ close to $0.5$ for all our simulations, approaching $0.4$ only for some of the most compact configurations. One of the possible reasons for the deviation from a simple proportionality is that Eq.~\ref{eq:RAlf} assumes the flat-spacetime dipole scaling $B = \mu r^{-3}$, violated for small radii in Schwarzschild spacetime. Nonetheless, our results show that the scaling $ R_{\rm m} \propto R_{\rm A}$ survives reasonably well across different detailed system configurations, with the standard power-law of $1/7$ in Eq.~\ref{eq:RAlf} outperforming $1/10$ proposed by \citet{Kulkarni2013}. The values of $\epsilon$ rise only slightly with the mass accretion rate, which is in agreement with the calculations of \citet{2017MNRAS.470.2799C}.

\subsection{Mass inflow and outflow}

\begin{figure*}
    \centering
    \includegraphics[width=1\linewidth]{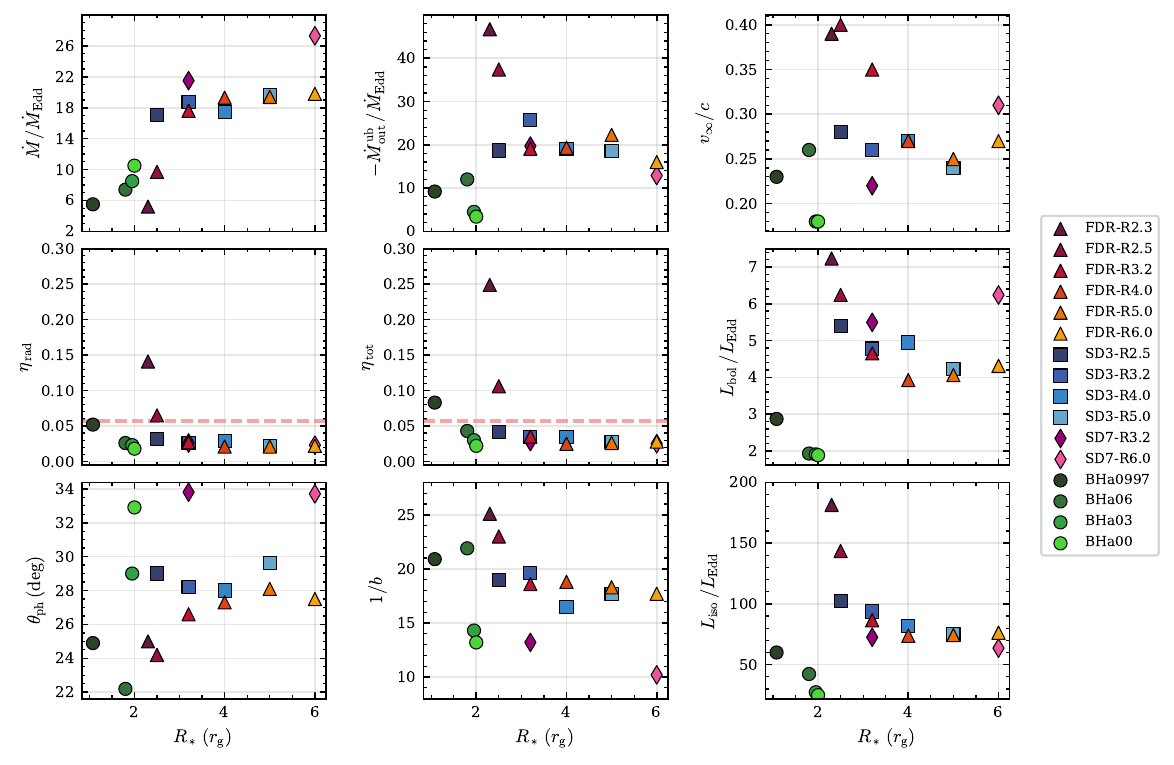}
    \caption{Parameters extracted from our GRRMHD simulations. Top row: net accretion rate $\dot{M}$, unbound outflow rate $\dot{M}^{\rm ub}_{\rm out}$ (both in units of $\dot{M}_{\rm Edd}$), and the upper limit on the terminal velocity of the unbound outflowing gas $v_\infty/c$. 
    Middle row: The radiation and total efficiencies $\eta_{\rm rad}$ and $\eta_{\rm tot}$, and the bolometric luminosity $L_{\rm bol}$. 
    Bottom row: The funnel opening angle $\theta_{\rm ph}$, 
    inverse beaming factor $1/b$, and the apparent luminosity $L_{\rm iso}/L_{\rm Edd}$, all as functions of the compact object radius. For the compact star group, the radius is the stellar radius $R_{*}$, and for BH group, it is the event horizon radius, both in units of $r_{\rm g}$. Each simulation model is represented by a different marker and colour. The dashed horizontal line in the efficiency panels indicates the thin disc accretion efficiency for a zero-spin compact object, $\eta_0 = 0.057$. }
    \label{fig:parameters}
\end{figure*}

Fig.~\ref{fig:momentum} shows the geometry of the accretion flow and outflow for a selected simulation from each category. 
The photospheric geometry varies between the models. From \texttt{FDR-R3.2} to \texttt{SD7-R3.2}, the photospheric funnel becomes progressively wider, and the unbound region becomes narrower. Among the models shown, \texttt{BHa0997} has the narrowest optically thin funnel in this selection, which is related to the presence of strong optically thick outflows. The photosphere opening angle for all simulations is compared in the bottom-left panel of Fig.~\ref{fig:parameters}.

Mass accretion rate for all simulations is presented in Fig.~\ref{fig:parameters} (top left panel). Within each family of models, the mass accretion rate increases with the star (or event horizon) radius. The trend is opposite for the outflowing component $-\dot{M}^{\rm ub}_{\rm out}$, see Fig.~\ref{fig:parameters} (top middle panel). A proxy for the mass inflow rate $\dot{M}_{\rm in} = \dot{M} - \dot{M}^{\rm ub}_{\rm out}$ is much more stable across setups, with $\dot{M}_{\rm in} \sim 15 \Medd$ for BHs and $\dot{M}_{\rm in} \sim 40 \Medd$ for stars. This illustrates how outflows reduce the net accretion rate and become more important in case of more compact configurations and for the stars, where the matter can not be advected through the event horizon.

\begin{figure}
    \centering
    \includegraphics[width=\linewidth]{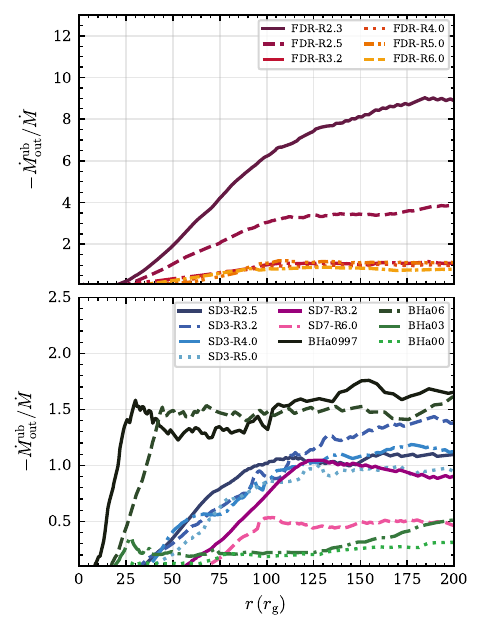}
    \caption{The radial profile of outflow rate of the unbound gas for all simulations normalized by its respective  mass accretion rate  (column $\Mdot$ in Table~\ref{tab:sims}). Each colour and line style indicates one simulation.}

    \label{fig:mdotout}
\end{figure}

While the net mass accretion rate $\Mdot$ is approximately independent of radius in a quasi-stationary simulation, the amount of inflowing and outflowing gas differs with radius. Hence, in Fig.~\ref{fig:mdotout} we show the radial profile of the unbound outflow rate $\dot{M}^{\rm ub}_{\rm out}$ for all simulations. We normalize the $\dot{M}^{\rm ub}_{\rm out}$ values by the respective net $\Mdot$ in each simulation to be able to compare the relative outflow strength. The \texttt{FDR} sequence is shown in a separate panel because the values for the strongly outflow-dominated simulations \texttt{FDR-R2.3} and \texttt{FDR-R2.5} are much higher than those of the other runs. In the star simulations, significant outflows develop above $r = 20\,r_g$ and increase with radius, see also the optically thick wind geometry in Fig.~\ref{fig:momentum}. Only for the case of the fast-spinning BHs $\dot{M}^{\rm ub}_{\rm out}$ rises steeply in the innermost region, indicating the compact zone of the Blandford--Znajek jet launching \citep{blandfordElectromagneticExtractionEnergy1977}.

\subsection{Outflow velocities}
\label{subsec:outflow_velocity}

To compare our simulations with the ultra-fast outflows observed in ULXs, we estimate the terminal velocity of the wind ${v_\infty}$ in the region around the photosphere. Assuming that the thermal, magnetic and trapped radiation energy is fully converted into kinetic energy and comparing $\xi$ (Eq.~\ref{eq:mu}) to the specific kinetic energy of the outflow, we give an upper limit for the velocity at infinity
\begin{equation}\label{eq:vinf}
    \frac{v_\infty}{c} \lesssim \sqrt{1 - \frac{1}{(\xi+1)^2}} \ .
\end{equation}
We measure this velocity just below the photosphere in between $r = 100\,\rg$ and $300\,\rg$, in the area where $1 \leq \tau < 2$, where the gas is unbound ($\xi > 0$) and outflowing ($u^r > 0$), see the system geometry in Fig.~\ref{fig:momentum}. We report the average $v_\infty$, weighted by $\rho u^r$, in Fig.~\ref{fig:parameters} (top right panel).

The most compact stars with strongest surface magnetic fields (first three models in the \texttt{FDR} sequence) launch the fastest winds, with $v_\infty \approx 0.35$--$0.40\,c$. The terminal velocities are comparable to the ultra-fast outflows of $\sim 0.2$--$0.25\,c$ detected in ULXs \citep{2016Natur.533...64P,2016ApJ...826L..26W,2018MNRAS.479.3978K} for a range of values of compactness and magnetic field configurations. In case of BHs, a large spin is necessary for $v_\infty > 0.2\,c$.

\subsection{Luminosities}

\begin{figure*}
    \centering
    \includegraphics[width=1\linewidth]{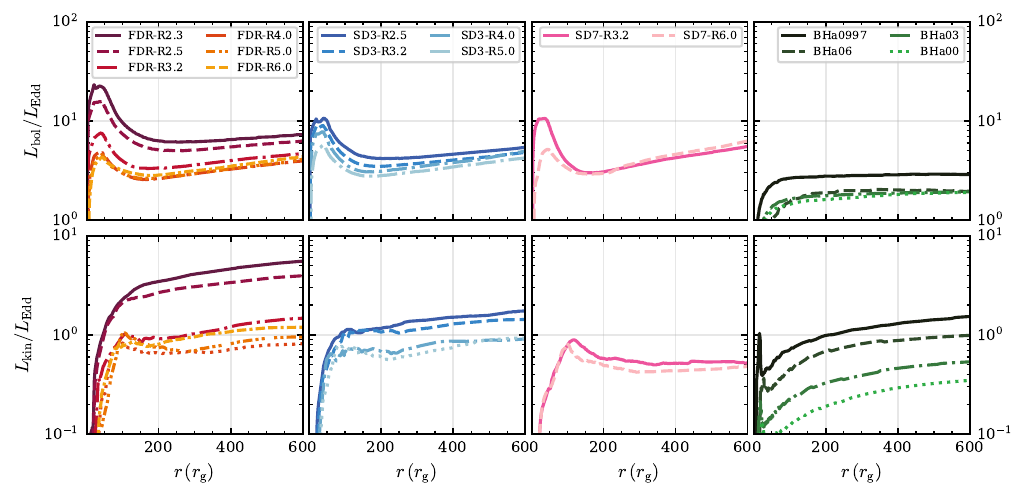}
    \caption{Radial profiles of radiation luminosity $L_{\rm bol}/\Ledd$ (first row) and kinetic luminosity $L_{\rm kin}/\Ledd$ (second row) as a function of radius $r/\rg$ for all simulation models (\texttt{FDR}, \texttt{SD3}, \texttt{SD7}, and \texttt{BH}). Each colour and linestyle represents an individual simulation within a given model. The increase in the innermost regions of the BH runs corresponds to the fast orbital motion, since the definition (Eq.~\ref{eq:kinenergylum}) includes orbital kinetic energy as well. We therefore use $\xi > 0.03$ in Eq.~\ref{eq:kinenergylum} to exclude the marginally unbound, rotating disc atmosphere near the BH, whose orbital kinetic energy otherwise dominates $L_{\rm kin}$ within $\sim 100\,\rg$.} 
    \label{fig:luminosity}
\end{figure*}

The bolometric luminosities (Eq.~\ref{eq:lbol}), as well as the maximum isotropic (apparent) luminosities (Eq.~\ref{eq:liso}) measured in the optically thin part of the domain, are given in Fig.~\ref{fig:parameters}. Both luminosities increase with compactness for stars and BHs alike. However, there is a clear separation between BHs and star systems with the latter producing higher luminosity values. The intrinsic radiative power of BH systems grows with spin and compactness, but it does not exceed $3\,\Ledd$ for any of the considered cases. On the other hand, in all star simulations the bolometric luminosity is approximately between 4 and 7 $\Ledd$. The lower radiative luminosity of the BH simulations is mostly a consequence of generally higher mass accretion rates to which the star simulations settle. However, at least for the most compact \texttt{FDR} stars, the difference is rather related to the radiative efficiency, see Section~\ref{sec:efficiencies}. A clear trend of more compact stars producing more intrinsic radiative power is broken only by the $\texttt{SD7-R6.0}$ model, which is the least compact, has by far the largest magnetic dipole moment and magnetospheric radius (see Fig.~\ref{fig:alfven}).

In Fig.~\ref{fig:luminosity}, the radial dependence of measured radiative luminosity $L_{\rm bol}$ and kinetic luminosity of the outflowing plasma $L_{\rm kin}$ is presented. Radiation emerges at the partially reflective star surface $R_*$ and its total power grows throughout the inner region of strong emission, while the outflow power becomes appreciable above the magnetosphere. The kinetic power of the outflow then increases with radius as more matter becomes entrained in the outflowing wind (see also Fig.~\ref{fig:mdotout}) and as the outflow accelerates toward the terminal velocity $v_{\infty}$ (see Section \ref{subsec:outflow_velocity}) through magnetic and radiative energy conversion into kinetic energy. This energy conversion results in decreasing $L_{\rm bol}$ approximately in the region of $50-200\, \rg$.  At larger radii, the luminosity remains approximately constant out to $r = 600\,\rg$, with values in the range $\sim 3$-$7\,\Ledd$. There is a slight increase visible at the outermost radii in some of the star simulations, which can be attributed to the change of the photosphere shape due to lower opacities in the outer region and a related release of the radiation carried out with the initially optically thick winds. Similarly as the radiative power, the kinetic power generally grows with compactness for both stars and BHs.

Fig.~\ref{fig:luminosity} also shows that, in the magnetized star cases, the radiative and kinetic luminosities have a different origin; while most of the radiative luminosity originates in the innermost regions around the central star, the outflows are gradually increasing throughout the domain. The kinetic luminosity alone does not clearly distinguish the star and BH systems (the peak of $L_{\mathrm{kin}}$ in the innermost regions of the BH simulations is attributed to the orbital kinetic energy, rather than to the outflows). In contrast, the strong emission from the innermost region with a local maximum of bolometric luminosity is characteristic of the compact star simulations and may provide a clearer indication of the presence of the stellar surface and magnetosphere.

\subsection{Efficiencies}
\label{sec:efficiencies}

The radiative efficiency in all our simulations, with the exception of the two most compact \texttt{FDR} simulations, is below that of a Novikov-Thorne thin disc around a static BH, that is $\eta_0 = 0.057$ \citep{novikov_astrophysics_1973}, see the second row, left panel of Fig.~\ref{fig:parameters}. In the case of BH simulations, the reduction of efficiency is consistent with the mechanism of photon trapping and radiative advection in super-Eddington accretion flows \citep{Ohsuga2002,SadowskiNarayan2016}. Furthermore, the values of efficiency within the $0.02-0.05$ range are generally similar to those of a supercritical slim disc accretion with a mass accretion rate of a few $\Mdot_{\rm Edd}$ \citep{Madau2014}. Including the kinetic efficiency does not change the ordering, and only pushes the \texttt{BHa0997} simulation above the total efficiency of $\eta_0$ (second row, second column of Fig.~\ref{fig:parameters}). 

The compact star simulations other than the two most compact \texttt{FDR} models are also characterised by a low radiative efficiency within the $0.02-0.03$ range. This results from an interplay between photons being trapped in an optically thick flow, disc truncation by the magnetosphere, and energy release at the star surface. For the most compact star models both radiative and kinetic efficiencies are significantly elevated, and the two most compact \texttt{FDR} models exceed the efficiency of all BHs, including the fast-spinning case. This is in part because of the higher compactness exposing the deeper part of the gravitational potential. Additionally, star magnetic field channels the flow, facilitating the escape of photons from an accretion column, thereby increasing the fraction of accretion power released as radiation \citep{2015MNRAS.454.2539M}.

\subsection{Beaming}

\begin{figure*}
    \centering
    \includegraphics[width=\linewidth]{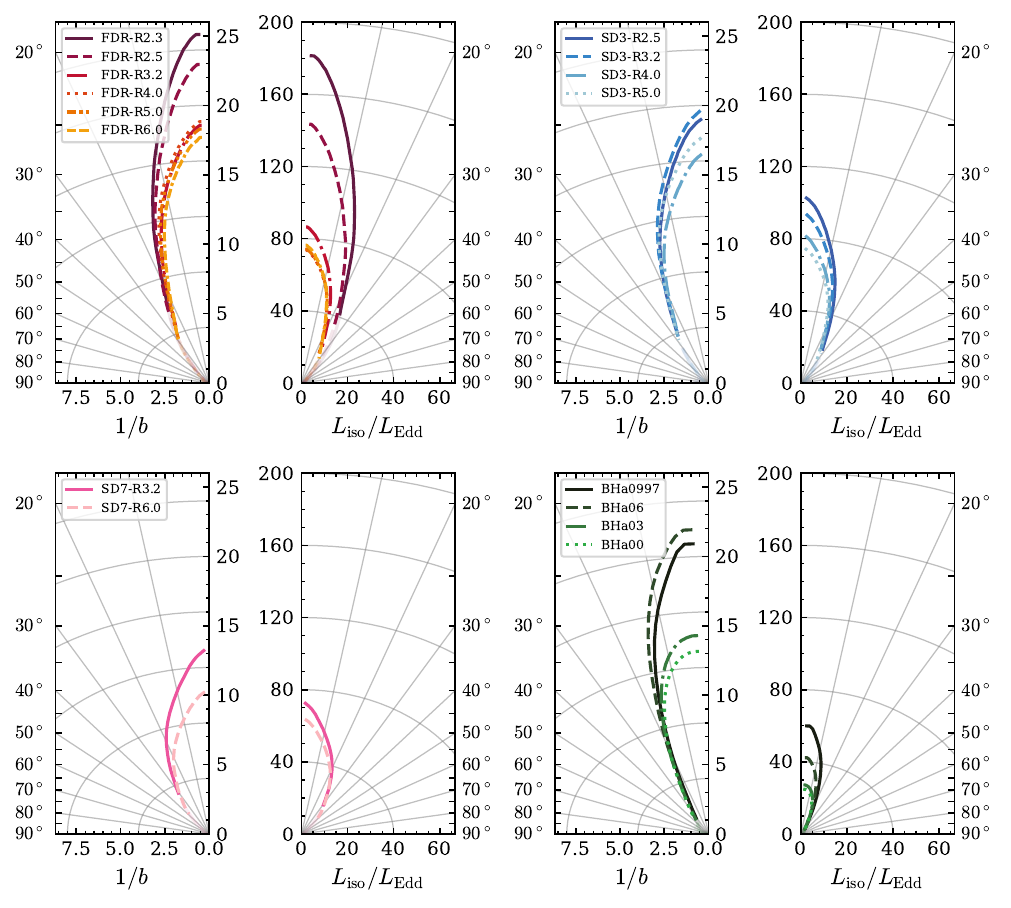}
    \caption{Inverse beaming factor $1/b$ and inferred isotropic luminosity $L_{\rm iso}/ L_{\rm Edd}$ in the left and right frames, respectively, as a function of viewing angle, measured from the pole. Each panel corresponds to one of the compact star sequences \texttt{FDR}, \texttt{SD3}, \texttt{SD7}, or to the \texttt{BH} group, and each colour and line style represents an individual simulation. The faded colour indicates regions that are optically thick and therefore do not contribute to the observed signal.}
    \label{fig:beaming}
\end{figure*}

In super-Eddington accretion systems radiation escapes through an optically thin conical funnel region along the axis of the rotation of the accretion disc. All of our simulations exhibit strong beaming -- radiation escaping along the polar funnel is collimated, and observers at low inclinations can detect luminosities strongly exceeding the intrinsic one, in qualitative agreement with the KLK model \citep{king_masses_2009, king2017, king2020}. We calculate the funnel opening angle $\theta_{\rm ph}$ by locating the optical depth $\tau_\theta = 1$ at each radius and averaging the corresponding polar angles over $100 \leq r/r_g < 300$. The results are presented in the bottom left panel of Fig.~\ref{fig:parameters}. For all simulations we find $22^\circ < \theta_{\rm ph} < 34^\circ$, similar as in super-Eddington BH accretion simulations of \citet{Narayan2017}. 

In Fig.~\ref{fig:beaming} we present a detailed study of the beaming factor $1/b$ and apparent luminosity $L_{\rm iso}$ dependence of the observer's viewing angle. Additionally, the peak values, corresponding to a viewing angle approaching $0^\circ$, are summarized in the bottom, middle and right panels of Fig.~\ref{fig:parameters}. Within star simulation families, beaming factor increases with compactness, reaching $1/b \approx 25$ for the most compact \texttt{FDR} model and a smallest value of $1/b \approx 10$ for \texttt{SD7-R6.0}, the least compact solution with the widest magnetosphere. Beaming in the BH systems is limited to a similar range of values, with two larger spin solutions resulting in $1/b \approx 21$ and two lower spin solutions yielding $1/b \approx 14$.

\section{discussion and conclusions}\label{sec:conc}

We have carried out the comparison of the super-Eddington accretion onto NSs, ultra-compact stars and light-mass BHs using GRRMHD simulations. By varying the radius and surface magnetic field strength of the compact object, we investigate how compactness and accretion geometry shape the radiative efficiency, outflow driving, beaming, and apparent luminosity, with the aim of placing these systems in the context of ULXs.

Within each stellar model series, the more compact objects generally accrete less mass but convert it into radiation and kinetic energy more efficiently and drive stronger unbound outflows. This causes a stronger geometrical beaming and higher apparent luminosities. The ultra-compact star simulation, \texttt{FDR-R2.3}, achieves a radiative efficiency of $\eta_{\rm rad} = 0.14$ with a beaming factor of $1/b \approx 25$, resulting in the maximal isotropic luminosity reaching $180\,\Ledd$ from the net mass accretion rate of $5.2 \dot{M}_{\rm Edd}$, corresponding to $5 \times 10^{40}$\,erg s$^{-1}$ for a $2.2\,M_\odot$ object, establishing it firmly as a bright ULX object. Nonetheless, all our simulations result in a maximum apparent luminosity well above $10^{39}$\,erg s$^{-1}$.

The interplay between compactness, magnetic field strength and their impact on the mass accretion rate, radiative efficiency, beaming and finally on the observable luminosity is complex. A larger magnetosphere suppresses beaming. For example, \texttt{SD7-R6.0}, with $R_{\rm m} \simeq 20\,\rg$, settles to the largest mass accretion rate, resulting in a large bolometric luminosity (while its radiative efficiency remains much lower than that of the most compact stars), but it indicates the widest funnel angle and the smallest beaming factor among all stellar models ($1/b \simeq 10$). At the same time \texttt{FDR-R6.0}, with the same stellar radius but a weaker magnetic field and smaller magnetosphere ($R_{\rm m} \simeq 10\,\rg$), reaches a lower mass accretion rate, similar efficiency, lower bolometric luminosity but a stronger beaming factor $1/b \simeq 18$. In the end, the two models end up with a roughly similar $L_{\rm iso}$. This highlights that the magnetospheric radius and the photosphere opening angle, shaped by the star-disc magnetic interaction, are considerable parameters alongside compactness and surface magnetic field, necessary to understand the system behavior.

The comparison between compact stars and BHs shows that BH models settle at lower mass accretion rate, similar to those of the most compact stars and less than a half of the measured accretion rate for the less compact stars. BHs radiative efficiency is low, but not lower than that of many of the less compact star solutions. Then, with a comparable beaming, BHs produce reduced luminosities, but given similar $\eta_{\rm rad} \times 1/b$ this is related primarily to the lower mass accretion rate. Thus, there is no dramatic (orders of magnitude) difference in luminosity between BH and star systems for the considered range of parameters, in contrast to the low mass accretion rate case \citep[e.g.,][]{Narayan2008}.

In this paper we also demonstrated that numerical GRRMHD simulations of magnetized stars support the standard Alfvén radius scaling for the magnetospheric radius $R_{\rm m} = \epsilon R_{\rm A}$, with $\epsilon \lesssim 0.50$, roughly consistent with the analytical thin disc estimations and with a mild dependence of $\epsilon$ on stellar compactness and mass accretion rate.

\koral{} uses the M1 closure scheme, which performs well for extended sources but produces unphysical beam interactions when radiation from distinct locations crosses. This is most relevant outside the accretion column and disc, where radiation directed towards the polar axis may interact with the reflected component from the polar boundary. A more accurate treatment of the beamed luminosity requires a radiative post-processing scheme relaxing the M1 approximation and resolving the frequency dependence of the emission \citep[e.g.,][]{Wielgus2022,2025ApJ...987...34D}. However, here we mainly provide a comparison between different models using the same numerical tools, and thus, since all models use the same numerical setup, the relative comparison between them is robust.


In order to connect our results to the observed properties of ULXs, we must address a difference in terminology: Our beaming factor defined by Eq.~\ref{eq:beam} should be interpreted with some care when compared to the \say{geometrical beaming} used in the ULX community. In the KLK and similar models, and in the Monte Carlo simulations \citep{2021MNRAS.501.2424M,2023MNRAS.518.5457M,2025MNRAS.543.1447K}, the central source is assumed to emit (quasi-)isotropically, and the amplification of the apparent luminosity arises from the reprocessing and collimation of this radiation by the optically thick walls of the wind cavity/accretion funnel. In our simulations, the emission emerging from the accretion columns and the central star is intrinsically anisotropic, directed preferentially along the rotational axis. Our beaming factor is therefore given by the combination of anisotropic emission and collimation by the optically thick outflow. A strong inclination dependence of the observed properties is indeed seen in ULXs \citep{2013MNRAS.435.1758S, 2015MNRAS.447.3243M}.

Consequently, our values of $1/b$ are not directly comparable to the amplification factors of the collimation models, nor to the constraints derived under the isotropic-source assumption, such as the argument that large pulsed fractions exclude strong beaming \citep{2021MNRAS.501.2424M}.
This distinction is also relevant for the observational constraints on beaming: several studies report only moderate beaming in individual ULXs from observed spectral and timing properties \citep{2015MNRAS.447.3243M, 2017MNRAS.467.3467P, 2018ApJ...856..128W}, optical counterparts \citep{2018ApJ...863..141B}, and from the large pulsed fractions of PULXs \citep{2021MNRAS.501.2424M}. In our simulations, the high apparent luminosities are not caused by extreme collimation (beaming factor $1/b > 25$) a large value of $L_{\rm iso}$ is based on the intrinsic luminosity, which is affected by compactness and the magnetic field strength of the star. Intrinsically anisotropic emission tied to the star may therefore reach ULX luminosities while remaining consistent with the moderate beaming and large pulsed fractions inferred observationally.

Notably, because the anisotropy in our models is intrinsic and connected to the stellar magnetic axis rather than produced by reprocessing in the outflow, it would co-rotate with the star in a misaligned configuration. Intrinsically anisotropic emission may therefore be compatible with both high apparent luminosities and the large pulsed fractions observed in PULXs, although confirming this would require 3D simulations of misaligned rotators and radiative post-processing, which we plan to focus on in the future.

\section*{Acknowledgments}
We thank the development team of \koral{} Olek S\k{a}dowski and David Abarca. FK acknowledges the program Vouchers for Universities in the Moravian-Silesian Region (registration number CZ.10.03.01/00/23\_042/0000390), within the project Accreting magnetized neutron stars. DL acknowledges the Czech Science Foundation (GAČR) project No. 25-16928O.  MW is supported by a Ramón y Cajal grant RYC2023-042988-I from the Spanish Ministry of Science and Innovation and acknowledges financial support from the Severo Ochoa grant CEX2021-001131-S funded by MCIN/AEI/ 10.13039/501100011033. MU acknowledges was supported by the MSK grant no. 04076/2024/RRC. This work was supported by the Ministry of Education, Youth and Sports of the Czech Republic through the e-INFRA CZ (ID:90254). We gratefully acknowledge Poland's high-performance Infrastructure PLGrid ACC Cyfronet AGH for providing computer facilities and support within computational grant no. PLG/2025/018773.
\section*{Software}\koral{} \citep{Sadowski2013, Sadowski2014}, Numpy \citep{numpy2011, numpy2020}, Scipy \citep{scipy2022}, Matplotlib \citep{matplotlib2007CSE}, and Pandas \citep{mckinney2010data}.
\section*{Data Availability}
The data underlying this article will be shared on reasonable request to the corresponding author.



\bibliographystyle{mnras}
\bibliography{bib} 




\bsp	
\label{lastpage}
\end{document}